# Exploration of increasing driver's trust in a semi-autonomous vehicle through real-time visualizations of collaborative driving dynamic

A. Koegel, C. Furet, T. Suzuki, Y. Klebanov, J. Hu, T. Kappeler, D. Okazaki, K. Matsui,
T. Hiraoka, K. Shimono, K. Nakano, K. Honma, M. Pennington

*DLX Design Lab, Institute of Industrial Science, The University of Tokyo*

## ABSTRACT

The Thinking Wave is an ongoing development of visualization concepts showing the real-time effort and confidence of semi-autonomous vehicle (AV) systems. Offering drivers access to this information can inform their decision making, and enable them to handle the situation accordingly and takeover when necessary. Two different visualizations have been designed: Concept one "Tidal" demonstrates the AV system's effort through intensified activity of a simple graphic which fluctuates in speed and frequency. Concept two "Tandem" displays the effort of the AV system as well as the handling dynamic and shared responsibility between the driver and the vehicle system. Working collaboratively with mobility research teams at the University of Tokyo, we are prototyping and refining the Thinking Wave and its embodiments as we work towards building a testable version integrated into a driving simulator. The development of the thinking wave aims to calibrate trust by increasing the driver's knowledge and understanding of vehicle handling capacity. By enabling transparent communication of the AV system's capacity, we hope to empower AV-skeptic drivers and keep over-trusting drivers on alert in the case of an emergency takeover situation, in order to create a safer autonomous driving experience.

## INTRODUCTION

As innovations in mobility progressively emphasize autonomy, establishing trust between the driver and vehicle will be a key factor in providing a safe driving experience. Over-trusting a vehicle system's capacity to drive and handle certain situations can lead to hazardous accidents. Under-trusting AVs in simple scenarios and forcefully taking the wheel could also increase the risk of accidents. In either case, these human trust factors give way to errors that could be mitigated with proper development of trust.

In addition to tackling trust between driver and vehicle on an individual level, it is important that the work is also carried out on a social level. The impact of vehicle safety is not only of concern to drivers but extends to other road users, including but not limited to in-vehicle passengers, surrounding vehicles, cyclists, and pedestrians. Legislative measures exist to protect all road users but these have proven to be obstacles for traditional original equipment manufacturers (OEMs) when deploying new innovations. Recently, Honda successfully pushed for an amendment to regulations enabling its own level 3 autonomous vehicles to operate under conditional autonomy in the streets of Japan (1).

Currently, the pioneers of semi-autonomous vehicle systems (operating at level 2) such as the Nissan ProPilot 2 and Tesla Autopilot convey autonomous control through a display which shows an illustration of the car in relation to other vehicles and obstacles on a road (2). This is displayed on the dashboard or on an adjacent monitor. The visualization seems almost video-game like, as drivers look through what feels like a small digital window. Other systems such as the Cadillac Super Cruise, build on such graphics by displaying a visual cue of hands on a steering wheel to suggest the driver takes over handling. While these systems may be sufficient for the current state of low/semi autonomous vehicles, they do not foster collaboration, hindering the ability to establish trust between driver and vehicle. Upcoming models like the BMW iNext strive for a level 3 autonomy where the responsibility can be shared by driver and vehicle with more flexibility (3). Still, the communication of vehicle capabilities is limited in many existing autonomous driving systems. We believe that as a driver, a notification of the level of autonomy while on the road is trivial. Instead of vaguely quantifying autonomy, empowering the driver so that they can acknowledge the capability of the vehicle in real-time scenarios is more effective in improving their handling ability.

How might we increase a driver's trust in a semi-autonomous vehicle by visualizing the collaborative driving process in real-time?

The Thinking Wave is a tool which visualizes the vehicle system's effort while tackling the current scenario. It is specifically designed for crossing the threshold from manual to level 1, 2 and 3 autonomy. As drivers learn to read the indicated confidence and handling capacity of the vehicle, they will feel more empowered to control and manage it. This knowledge will also help the driver understand the responsibility they have in their present scenario, and the level of attention that must be maintained. By calibrating expectations for collaboration we hope to increase trust in order to facilitate a safe driving and riding experience.

## CONCEPTS

Working with professors and mobility experts at the Institute of Industrial Science at the University of Tokyo, we designed two distinct visualization concepts for the Thinking Wave.

Concept one, named "Tidal", displays a wave that indicates the level of complication in the vehicle system's processing, suggesting when the driver should adjust control or takeover. It fluctuates in intensity (amplitude, frequency and speed) to reflect the effort of the vehicle system. To ensure it would be easy to read, we took inspiration from existing, familiar visualizations of biometric data such as human electrocardiograms (ECGs). Similar to how a human's increased heart-rate under stressful conditions would be translated, the vehicle would demonstrate intensity through multiple high-amplitude waves close to one another when going through a difficult handling situation. For example, in the instance of driving through heavy rain, visibility is lowered and the vehicle's sensors may be compromised in detecting its surroundings. As the system would struggle to adequately handle the situation, the wave would appear high, frequent and fast.

Although this concept is relatively straightforward, one of its limitations might be a lack of clarity in the visualization of system "effort." Through testing, we intend to understand whether this information is too shallow and whether it can be beneficial in communicating with and reassuring the driver.

Concept two, named "Tandem" is a multi-information layer visualization. It demonstrates the collaboration dynamic between vehicle and driver in addition to showing the vehicle's effort throughout various scenarios. Collaboration dynamic is displayed by a circle symbolizing the driver and a square symbolizing the vehicle system, both of which change in size, position and rotation speed. The leading controller within a scenario would be on the top (front) of the visual. The size of each shape shows the relative level of effort that it is putting into handling the situation. As for the vehicle system's thinking effort, the more complex the situation (regardless of who is in control), the faster it spins.

An example scenario would be a large front circle (driver) with a rapidly rotating, small square (AV system) while navigating through high-pedestrian in-city traffic. As the system does not feel confident handling such a complex and uncertain situation, the driver is expected to handle the situation.

Though Tandem is more abstract than Tidal, it has the potential to communicate richer and more nuanced information. Our intention through the human-computer interactions (HCI) is that they will aid the driver in determining the appropriate way to handle a situation. One possible limitation to consider while testing Tandem is its learnability factor. Though it shows potential for continuously educating drivers of evolving AV capabilities and calibrating trust, its interface is initially less intuitive than the Tidal interface. Therefore, we anticipate that test results may reflect negatively if learning time is inadequate.

Though the two concepts vary distinctly in visual representation and function, we created consistent elements to ensure the driver would be adequately equipped with proper knowledge and understanding of the situation before them. In conjunction with the central concept visuals, the perimeter of the windshield will light up to indicate where the driver should focus their attention as the scenarios evolve.

As previous research suggests, future-forward human-machine interface (HMI) of autonomous vehicles should diversify from being an exclusively visual experience. Rather it should expand to include multi-sensorial communication by incorporating audio and haptic feedback (4). This has the potential of broadening the HCI to include passengers as well as the driver.

We propose using haptic elements to adjust torque in correcting the driver's handling when possible and necessary, as well as highlighting any change in handling dynamic and takeover completion. Through audio, we suggest the use of two different sub-mediums: one in the form of alert jingles to acknowledge takeover, call to action (cta) and highlight specific occurrences; and the second as a voice to generate dialogue between driver and vehicle system. Commands, suggestions, navigational assistance and spoken notifications can be talked back to or ignored by the driver. These multi-sensory integrations will help emphasize the message delivered by the wave, in order to ensure that it is guiding the driver's attention to crucial areas and not simply providing a narration of the scenario and context. For example, rather than simply stating that the road is clear when the driver is making a turn, the system could encourage the driver to focus on a specific point to improve their handling. Based on feedback that we received, this type of passive communication is less risky in the case of a system failure as driver's wouldn't be reliant on the vehicle system to decipher road conditions. By keeping drivers engagement on the road and assisting through bringing awareness to danger, a safer experience can be maintained for the driver, passengers, and external stakeholders at level 3 autonomy.

## TEST DESIGN

The objective of prototyping these concepts is to test their ability and clarity in communicating vehicle system effort to the driver. Our hypothesis is that this would help to build trust between driver and vehicle and create a safer and more comfortable experience. The current prototype will be installed in the driving simulator at the University of Tokyo, Komaba Campus. Visualizations will be overlaid onto the existing infrastructure for users to experience the thinking wave while driving through a variety of scenarios. The intention is to learn how the wave influences driving behaviors of the participants in corresponding road conditions.

However, before we even reach this sophisticated level of testing, we will run pre-tests to measure just the perception and understanding of the two thinking waves by participants. This will be a simpler test in which participants will be shown a road condition and the corresponding wave on a monitor. They will be asked to respond what they think the wave means. Instead of giving participants a multiple choice answer set which could encourage guessing in unclear scenarios, their response will be written themselves. In this pre-test, there will be two test groups: one which is given an explanation of the waves beforehand and another that is not. By running this pre-test, we will be able to uncover whether the thinking wave concepts are understandable to begin with. From there, we will modify and refine the designs for the higher-fidelity test in the simulator. As the concepts are rather abstract, this will allow us to determine initial flaws in the thinking wave and quickly iterate even between each participant's test session. This sort of agility is a key driver in our design process as we will be able to quickly improve and deploy our prototypes.

The goal of the latter simulator test is to determine the participant driver's ability to decipher how they are to interact with the system in various scenarios and autonomous stages. The two waves will be projected onto the driving simulator windshield, effectively creating a windshield display. This will allow the driver to see both the wave and the "exterior scenario" played by the simulator. Speakers and haptic feedback will be linked to the wave interaction, creating a multi-sensory experience. An eye tracker and force-sensing resistors (FSR) integrated into the steering wheel, will also be installed, offering opportunities for data collection during testing. The test will consist of three distinct driving experiences.

A. The driver navigates through a path with no wave visualization
B. The driver navigates through the path with the Tidal wave visualization overlaid
C. The driver navigates through the path with the Tandem wave visualization overlaid

In both cases, before the visualization appears, participants will be shown a short demonstration video outlining how the concept works. After the drive, they will be asked several questions about the experience and their understanding of the wave.

It will be important to understand which of the two wave visualizations resonate most with the participants, as well as understanding the effect that they will have on their driving performance. Having access to more nuanced sensor data will help ensure that our findings are accurate: while the driver may indicate a preference for one concept, their performance may reflect otherwise. Incomplete data would risk undermining the longer-term potential of the concepts, particularly with the tandem wave, which has the potential of delivering deeper levels of information but might be more difficult to read initially.

In addition to the surveys, data will be collected in the following forms:
- Eye-tracking through the use of smart-eye and smart-recorder. This will help us pinpoint the driver's focal point throughout different scenarios to understand where their attention lies, and whether they are either sufficiently or overly focused on the wave.
- FSR sensing will give us an estimation of the driver's sense of trust in the system according to their hand placement (on/off) and grip strength on the steering wheel.
- A behavioral handling assessment will be conducted for each scenario segment within the overall narrative. Participants will be scored on how well they handled each scene (on a scale of 1-5). The total score from these quantified metrics will help us understand the user's interaction with the wave and potentially its effectiveness. Also, each scenario fragment's score uncovers the weaknesses of specific concept elements. This could determine what areas we should focus on for further refinement. Through their handling, we will be able to determine the driver's ability to comprehend the thinking wave.

## FUTURE EXPLORATION

Currently, AV innovation primarily focuses on improving autonomy, accuracy and handling capacity. A crucial consideration that is seemingly overlooked is the driver's experience when the vehicle is in autonomous mode, particularly with regard to takeover situations.

It is important for us as designers to carefully consider what drivers are doing when the vehicle is fully autonomous, and how we can keep them passively engaged so they are ready to take over if necessary. This is particularly relevant when designing for near future automation, levels 1-4. How might we create an enjoyable and relaxing experience for drivers who are in the driver's seat but not driving, while still keeping them alert and prepared?

The thinking wave concept shows potential in helping drivers not only calibrate their expectations but also establish trust in AVs. We see it as a tool that will evolve alongside both the system technology and driver to build a safer and more effective AV future. If the thinking wave can learn the driver's handling tendencies, it could help the vehicle system adapt and control according to preference. Furthermore, with software updates and enhancements in autonomous systems, the wave's confidence visualization can be altered to help drivers recalibrate their trust. This concept has the potential to educate drivers while enabling innovation and adoption of new technology.

## APPENDICES

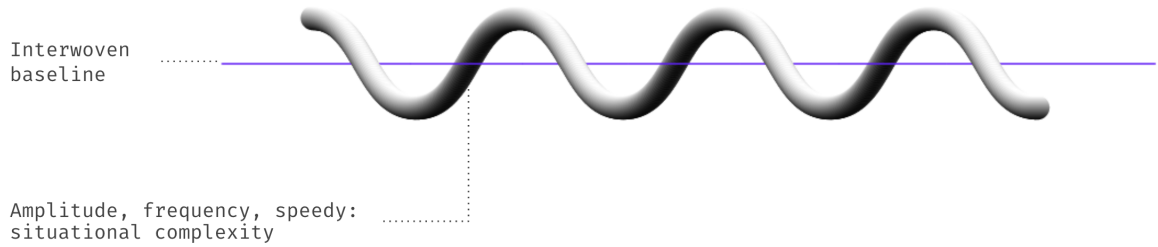

*1 - Annotated diagram of "tidal" visualization*

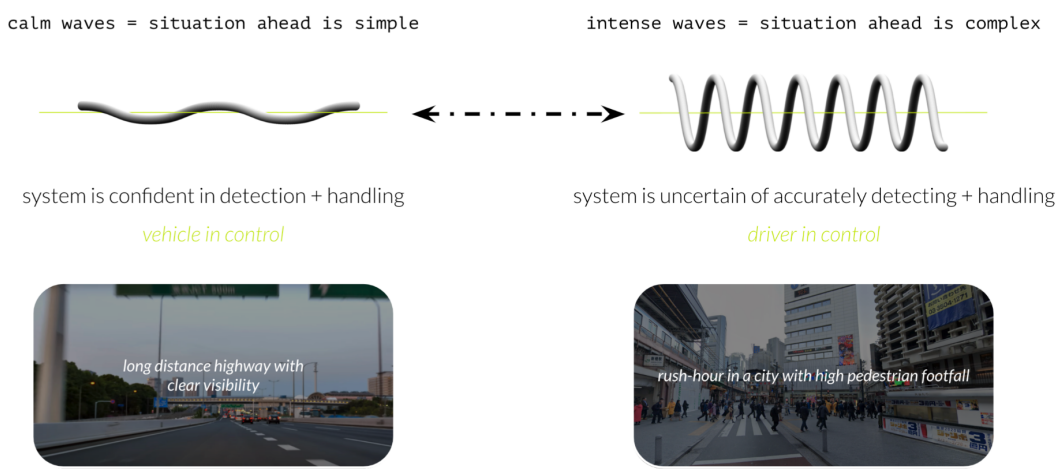

*2 - Explanations and example display scenarios of "tidal" variations in "calm" [autonomous] and "intense" [manual] state*

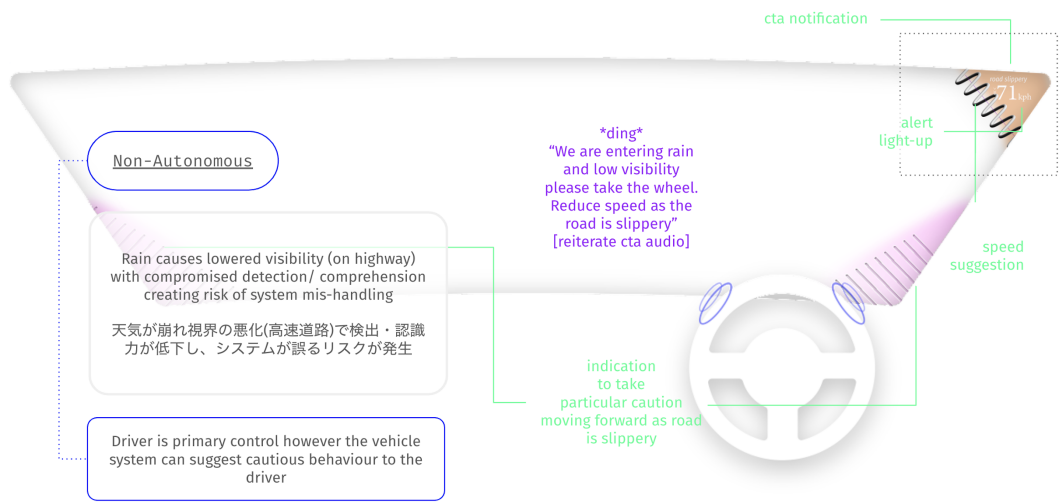

*3 - "Tidal" concept in an example scenario in conjunction with other reinforcement sensory elements (wave displayed in top right corner of windshield).*

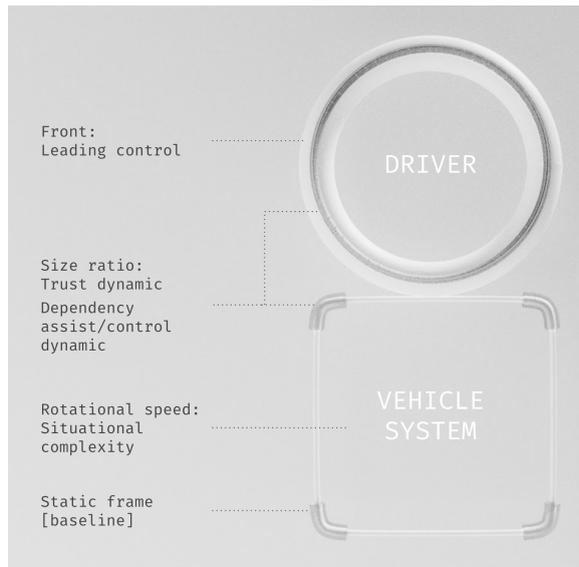

*4 - Annotated diagram of "tandem" visualization*

icon sizes equal = equal control
rotational speed slow = neutral complexity

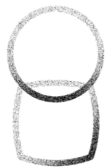

driver and system are working together
*collaborative control*

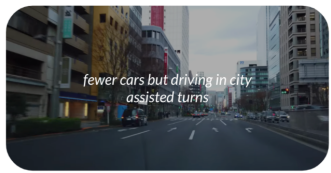

fewer cars but driving in city assisted turns

large, front circle = driver is primary controller
small, fast square = vehicle system overloaded

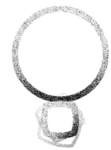

vehicle system is unable to sense/control with confidence
*driver in control*

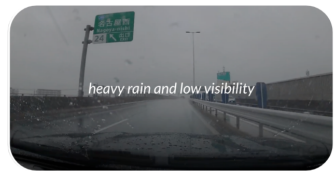

heavy rain and low visibility

square is main = vehicle system in control
circle inside = driver's hands off

level three autonomous mode where driver can take hands off the wheel
*vehicle in control*

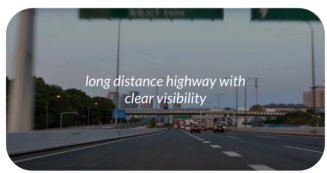

long distance highway with clear visibility

dominant square = vehicle system is primary control

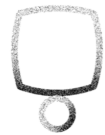

level 2 autonomous mode where driver should still keep hands on wheel
*vehicle dominant control, driver ready to assist/take-over*

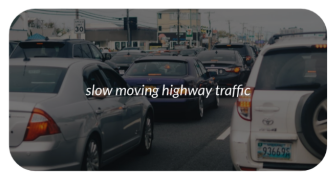

slow moving highway traffic

*5 - Explanations and example display scenarios of "tandem" variations in equal collaborative state, fully manual state, autonomous state and semi-autonomous with driver assisting state*

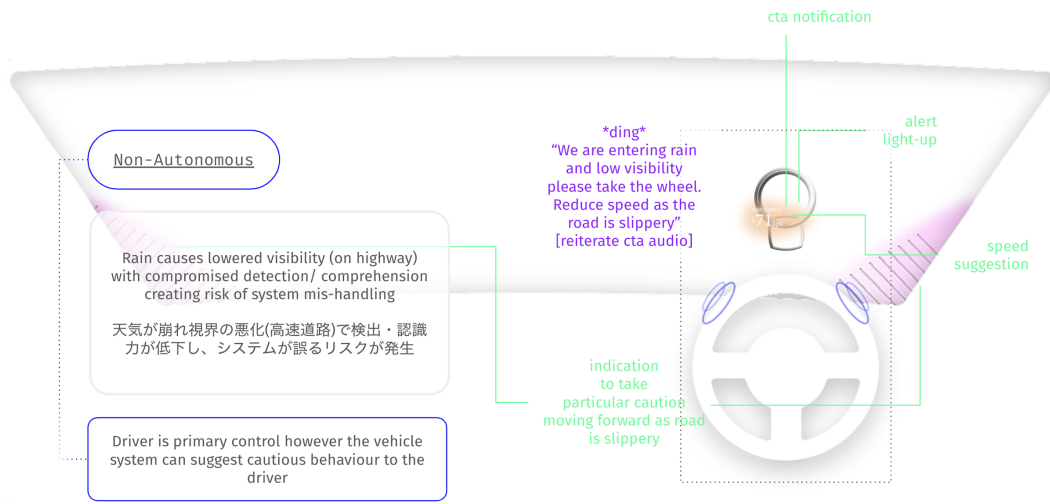

*7 – "Tandem" concept in an example scenario in conjunction with other reinforcement sensory elements (wave displayed on windshield above steering wheel).*

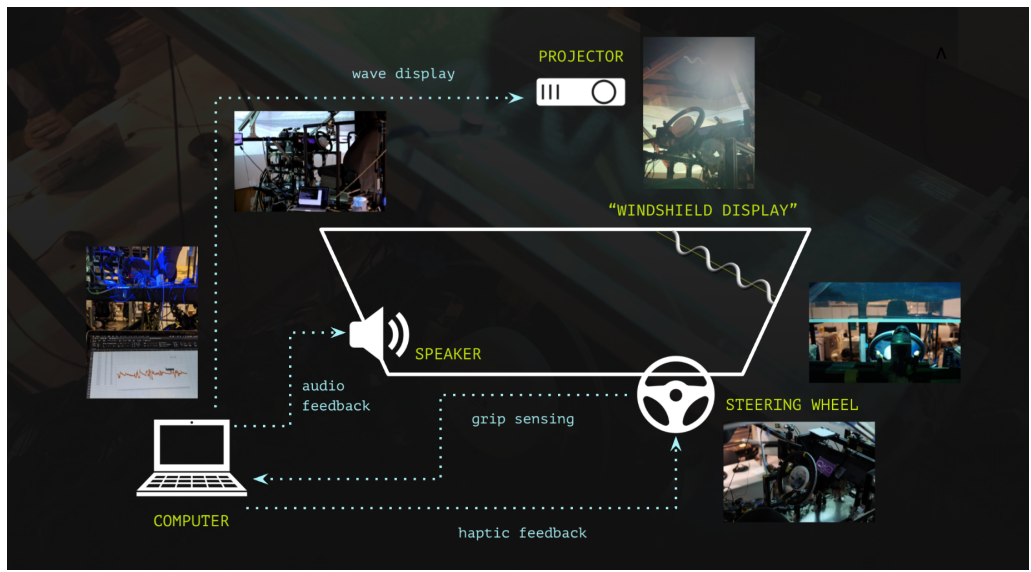

*8 – Diagram of the thinking wave prototype integrated into the driving simulator*

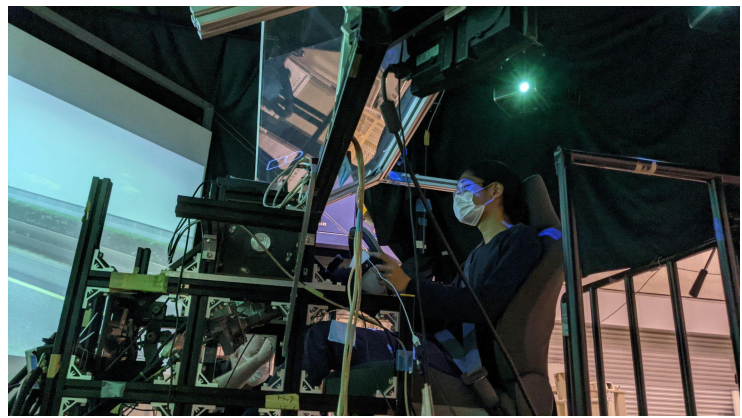

*9 – Driving simulator with tandem wave prototype displayed on the windshield*

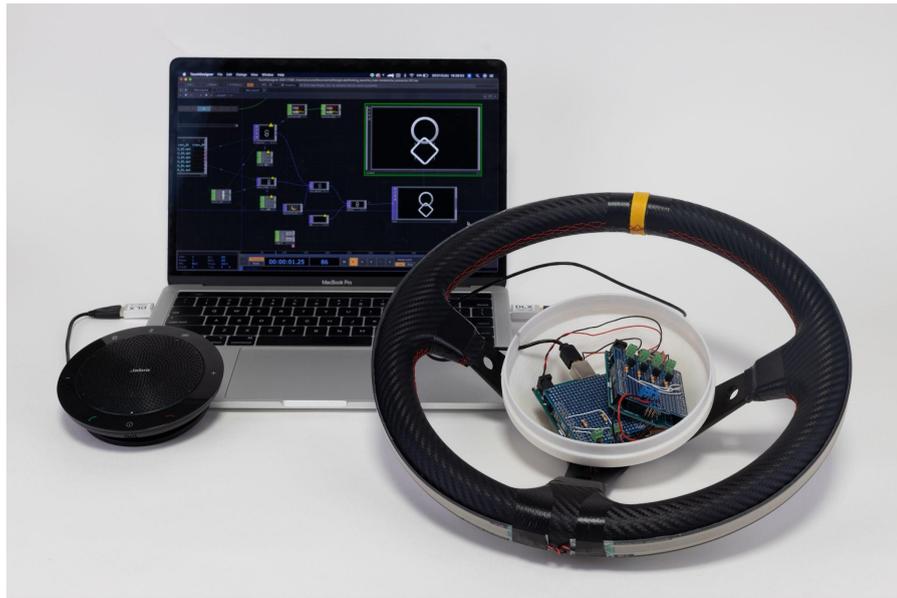

*10 - Steering wheel hardware components integrating FSR touch/grip sensors and haptic notification vibration motors. Touch-designer software running tandem variation animations.*

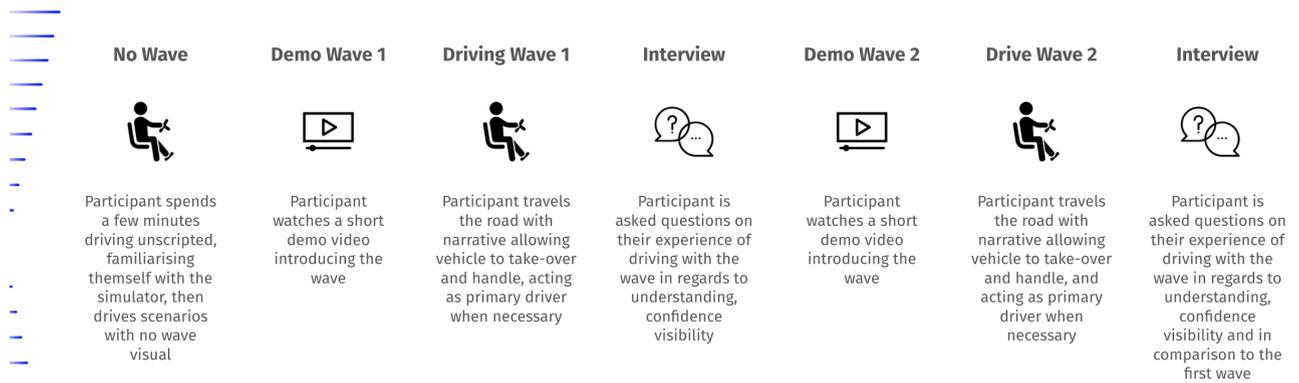

*11 - Flow of the handling behavior test to take place in the simulator*